\documentclass{PoS}

\usepackage{amsmath}
\usepackage{graphicx,color}

\newcommand{\id}{{\bf 1}}
\newcommand{\arxiv}[1]{{arXiv:#1}}

\newcommand{\A}{X}

\title{New fermion discretizations and their applications}

\ShortTitle{New fermion discretizations}

\author{\speaker{Tatsuhiro Misumi}\thanks{The author is supported by
        the JSPS Postdoctoral Fellowship for Research Abroad (No.24-8). }\\
        Brookhaven National Laboratory, Upton, NY 11973, USA\\ 
        E-mail: \email{tmisumi@bnl.gov}}

\abstract{We review the recent progress in new lattice fermion formulations. We focus on the following three types which have possibility of improving lattice simulations. (1) Flavored-mass fermions are a generalization of Wilson fermions with species-splitting mass terms. In particular, staggered-Wilson fermions initiated by Adams have possibilities of reducing numerical costs in overlap fermions and the influence of taste-breaking in staggered fermions. (2) Central-branch Wilson fermions, in which additive mass renormalization is forbidden by extra axial symmetry, could enable us to perform Wilson-fermion lattice QCD without fine-tuning. (3) Minimally doubled fermions, which reduce the number of species by species-dependent chemical potential terms, realizes a ultra-local chiral fermion at the price of hypercubic symmetry. These setups reveal unknown aspects of lattice fermions, and we obtain a deeper understanding of lattice field theory.}

\FullConference{The 30 International Symposium on Lattice Field Theory - Lattice 2012,\\
		June 24-29, 2012\\
		Cairns, Australia}

\begin{document}

\section{Introduction}
\label{sec:Intro}

Non-perturbative lattice calculations have contributed to a broad range of 
topics in particle and nuclear physics.
It becomes more and more important in the era of new physics in LHC.
However, practical problems in lattice fermions still prevent us from performing simulations efficiently. By now several fermion constructions have been developed, although all of them have their individual shortcomings. To perform more efficient and precise lattice simulations and to make progress in the frontier of high energy physics, it is of primary importance to develop lattice fermions with less numerical costs, less discretization errors, less tastes and better chiral property. 
Although effective ways of improving have been proposed and are used widely in the calculations, an interesting goal is to construct new fermion formulations free from the shortcomings. The research in this course will also give us a new insight into lattice field theory and feedback for the existing simulations.

In this paper we review the recent progress on new fermion setups and discuss their applications in lattice simulations. In Sec.~\ref{sec:FM} we discuss generalization of Wilson fermions by using species-dependent mass terms, which we call  ``flavored-mass" \cite{Adams, Hoel, CKM1}. In this study we will find new versions of Wilson, domain-wall and overlap fermions, some of which successfully have better dispersion relation and better eigenvalue spectrum \cite{DK}. In particular staggered-Wilson (overlap) fermions \cite{Adams, GS} have the possibility of reducing overlap computational costs \cite{PdF, CKM2} and reducing staggered taste breaking effectively \cite{steve, MNKO}. We secondly discuss a different use of Wilson-type fermions. In Sec.~\ref{sec:CB} we will show that the Wilson fermion without non-hopping terms has an extra $U(1)$ symmetry, which forbids the additive mass renormalization \cite{CKM2, Rev}. This axial symmetry is spontaneously broken with the associated NG boson emerging. By combining this idea to the flavored mass, we in principle obtain a two-flavor central-branch fermion which could suit two-flavor lattice QCD. In Sec.~\ref{sec:MD} we discuss another way of keeping chiral symmetry and reducing the number of flavors in the lattice fermion. ``Minimal-doubling setup" reduces the species to two with keeping one exact chiral symmetry at the price of hypercubic symmetry \cite{KW, CB, CM}. To perform lattice QCD with this fermion, we need to tune the several parameters to restore Lorentz invariance in the continuum limit \cite{Bed, Cap}. Recently \cite{misumi} introduced a new view that this setup can be interpreted as a lattice fermion with species-dependent chemical potential terms. Application of this fermion to in-medium QCD is discussed in \cite{MKO}. In Sec.~\ref{sec:S} we discuss the perspective on these formulations.


\section{Flavored mass and generalized Wilson fermions}
\label{sec:FM}

In this section we discuss generalizations of the Wilson fermion
from the viewpoint of species-splitting mass terms, which
we call ``flavored mass terms". 
It leads to new types of domain-wall and overlap fermions.
This course is initiated by Adams in \cite{Adams}, and the 
intensive researches have been done by now. 
This topic is roughly classified into two parts:
generalization of Wilson fermion based on naive fermions
and its generalization based on staggered fermions.
We begin with the former one and move to the latter.

\subsection{Flavored mass for naive fermions} 
 \label{subsec: FN}
 
\begin{figure}
\begin{center}
\includegraphics[bb=0 200 1024 550, clip, width=8cm]{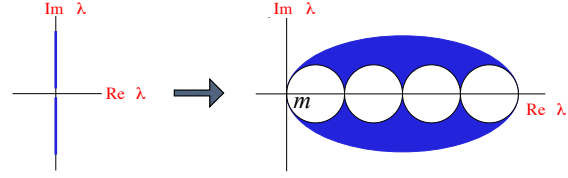} 
\end{center}
\caption{Free Wilson Dirac spectrum.
The degenerate spectrum of 16 species in naive fermions are 
split into five branches with 1, 4, 6, 4 and 1 flavors.}
\label{Wil}
\end{figure}

In this subsection we introduce flavored-mass terms for naive fermions 
as generalizations of the Wilson term \cite{CKM1}.
The Wilson fermion action is given by,
\begin{equation}
S_{\rm W} =\sum_{n,\mu}\bar{\psi}_{n}\gamma_{\mu}D_{\mu}\psi_{n}
\,+\, \sum_{n}m_{0}\bar{\psi}_{n}\psi_{n}
+\,r \sum_{n,\mu}\bar{\psi}_{n}(1-C_{\mu})\bar{\psi}_{n},
\label{WilS}
\end{equation}
where $D_{\mu}\equiv(T_{+\mu}-T_{-\mu})/2$, 
$C_{\mu}\equiv (T_{+\mu}+T_{-\mu})/2$ with $T_{\pm\mu}\psi_{n}=U_{n,\pm\mu}\psi_{n\pm\mu}$.
The free Dirac spectrum for the Wilson fermion is schematically depicted in Fig.~\ref{Wil}.
The degeneracy of 16 modes in naive fermions is lifted into 5 branches, 
to which 1, 4, 6, 4 and 1 flavors correspond.
We emphasize the three important properties of the Wilson fermion,
$\gamma_{5}$-hermiticity, hypercubic symmetry and the lattice laplacian form
$\sim a\int dx^4 \bar{\psi}_{x}\Delta\psi_{x} +O(a^{2})$.
These can be criterions for generalization.

Now we briefly look into flavor-chiral symmetry of naive and Wilson fermions by following \cite{Rev}. 
As well-known the massless naive action possesses $U(4)\times U(4)$, which is regarded as remnant of the continuum flavor-chiral symmetry group for 16 flavors.
This $U(4)\times U(4)$ is given by
\begin{equation}
\label{sym_naive_all}
\psi_n\rightarrow\exp\Big[i \sum _\A \left(\theta _\A^{(+)}\Gamma^{(+)}_\A+\theta _\A^{(-)}\Gamma^{(-)}_\A\right)\Big]\psi_n\,,\,\,\,\,\,\,\,\,
\bar{\psi}_n\rightarrow\bar{\psi}_n\exp\Big[i\sum _\A\left(-\theta _\A^{(+)}\Gamma^{(+)}_\A+\theta _\A^{(-)}\Gamma^{(-)}_\A\right)\Big]\,.
\end{equation}
Here, $\Gamma^{(+)}_\A$ and $\Gamma^{(-)}_\A$ are site-dependent $4\times 4$ matrices:
\begin{eqnarray}
\label{sym_t+}
\Gamma^{(+)}_\A&= &\left\{\mathbf{1}_4\,,\,\, (-1)^{n_1+\ldots+ n_4}\gamma_5\,,\,\,(-1)^{\check{n}_\mu}\gamma_\mu\,,\,\,(-1)^{n_\mu}i \gamma_\mu\gamma_5
\,,\,\,(-1)^{n_{\mu,\nu}}\frac{i \,[\gamma_\mu\,,\gamma_\nu]}{2}\right\}\,,\\
\label{sym_t-}
\Gamma^{(-)}_\A&= &\left\{(-1)^{n_1+\ldots+ n_4}\mathbf{1}_4\,,\,\, \gamma_5\,,\,\,(-1)^{n_\mu}\gamma_\mu\,,\,\,(-1)^{\check{n}_\mu}i \gamma_\mu\gamma_5
\,,\,\,(-1)^{\check{n}_{\mu,\nu}}\frac{i \,[\gamma_\mu\,,\gamma_\nu]}{2}\right\}\,,
\end{eqnarray}
where $\check{n}_\mu=\sum _{\rho\neq\mu}n_{\rho}$, $n_{\mu,\nu}=n_\mu+n_\nu$ and $\check{n}_{\mu,\nu}=\sum_{\rho\neq\mu,\nu}n_\rho$. See \cite{Rev} for details. 
Quark condensate or quark mass break this $U(4)\times U(4)$ down 
to the $U(4)$ vector subgroup $\Gamma^{(+)}_\A$.
We call $\Gamma^{(+)}_{\A}$ as vector-type group and $\Gamma^{(-)_\A}$ as 
axial-type group. 
In the presence of the Wilson term this $U(4)\times U(4)$ invariance is broken down 
to the $U(1)$ invariance under $\mathbf{1}_{4}$ in Eq.(\ref{sym_t+}).
This generator is vector-type, which means that the Wilson fermion loses all the axial(chiral) symmetry. 

Now we go on to the main theme ``flavored-mass terms".
In \cite{CKM1}, it was shown that there are four nontrivial types of flavored masses for naive fermions, 
which satisfy $\gamma_{5}$-hermiticity, possess the hypercubic symmetry 
and becomes covariant laplacian with proper mass shifts.
The four types are classified based on the number of transporters, 
where we name the 1-link case as vector (V), 2-link as tensor (T), 
3-link as axial-vector (A) and 4-link as pseudo-scalar (P),
\begin{equation}
 M_{\mathrm{V}} =  \sum_{\mu} C_\mu, \,\,\,\,\,\,\,
 M_{\mathrm{T}} =  \sum_{perm.}\sum_{sym.}C_\mu C_\nu ,\,\,\,\,\,\,\, 
 M_{\mathrm{A}} =  \sum_{perm.}\sum_{sym.}  \prod_{\nu} C_\nu,\,\,\,\,\,\,\,
 M_{\mathrm{P}} =  \sum_{sym.} \prod_{\mu=1}^4 C_\mu,
\label{F-mass}  
\end{equation}
where $\sum_{perm.}$ means summation over permutations of the
space-time indices.
$\sum_{perm.}$ and $\sum_{sym.}$ are defined as containing
factors, for example $1/4!$ for $M_{\mathrm{P}}$.
The vector type $M_{\mathrm{V}}$ is the usual Wilson term up to the mass shift,
and the Wilson term in (\ref{WilS}) is just given as $\sum_{n}\bar{\psi}_{n}(4-M_{\mathrm{V}})\psi_{n}$.
In the momentum space, they are transformed into $M_{\mathrm{V}}\to\cos p_{\mu}$,
$M_{\mathrm{T}}\to\cos p_{\mu} \cos p_{\nu}$, 
$M_{\mathrm{A}}\to\cos p_{\mu}\cos p_{\nu} \cos p_{\rho}$ 
and $M_{\mathrm{P}}\to\cos p_{1} \cos p_{2} \cos p_{3} \cos p_{4}$.
As shown in \cite{CKM1}, these flavored masses are also expressed in the spin-flavor representation
as $M_{\mathrm{P}} \sim ( {\bf 1} \otimes (\tau_{3}\otimes\tau_{3}\otimes\tau_{3}\otimes\tau_{3}))$.

\begin{figure}
\begin{center}
\includegraphics[bb=0 320 595 505, clip, width=13cm]{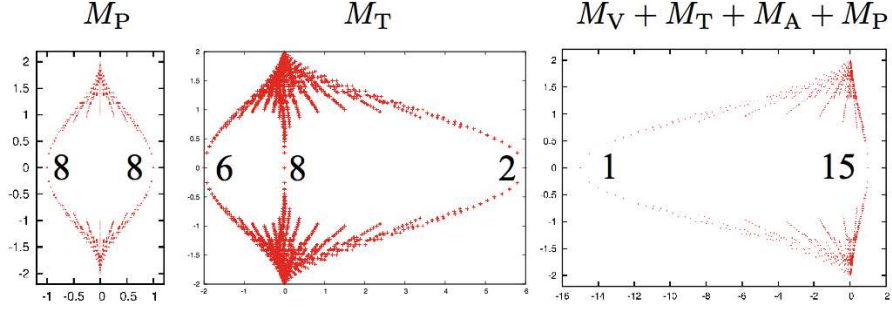} 
\end{center}
\caption{(Left) Dirac spectrum for $M_{\rm P}$ \cite{CKM1}, where we have two branches with 8 and 8 flavors.
(Center) Dirac spectrum for $M_{\rm T}$, where we have three branches with 6, 8 and 2 flavors.
(Right) Dirac spectrum for $M_{\rm V}+M_{\rm T}+M_{\rm A}+M_{\rm P}$ \cite{CKM1}, where we have
two branches with 1 and 15 flavors. }
\label{Mp}
\end{figure}

By introducing the flavored-mass terms $\bar{\psi}_{n}M_{F}\psi_{n}$ ({\it F}=V, T, A, P)
or their combinations into the naive action instead of the usual $M_{\rm V}$, 
we obtain generalized versions of Wilson fermions.
The species-splitting depends on the types or combinations.
$M_{\mathrm{P}}$ splits the 16 degenerate modes into two branches
with 8 and 8 flavors. The Dirac spectrum for $D_{\rm Naive}+M_{\mathrm{P}}$ 
is depicted in the left figure of Fig.~\ref{Mp}.
$M_{\mathrm{T}}$ splits them into three branches with 6, 8 and 2 flavors as depicted in the center figure of Fig.~\ref{Mp}.
$M_{\mathrm{A}}$ has the same species-splitting as the case of $M_{\mathrm{V}}$ (Wilson).
Combinations of these flavored-mass terms are also useful to split the species in a desirable form.
For example, the total sum of the flavored mass terms $M_{\rm V}+M_{\rm T}+M_{\rm A}+M_{\rm P}$
splits the 16 species into two branches with 1 and 15 flavors as shown in the right figure of Fig.~\ref{Mp}.
We note the hypercubic symmetry remains as long as we consider a combination of the forms 
in (\ref{F-mass}). 
The remaining flavor symmetry also depends on the types.
For example, $M_{\mathrm{P}}$ with the mass shift breaks 
the symmetries of naive fermions in Eqs.(\ref{sym_t+})(\ref{sym_t-}) into the subgroup
\begin{equation}
\label{sym_t+P}
\Gamma^{(+)}\,\,\to\,\,\left\{\mathbf{1}_4\,,\,\, (-1)^{n_1+\ldots+ n_4}\gamma_5
\,,\,\,(-1)^{n_{\mu,\nu}}\frac{i \,[\gamma_\mu\,,\gamma_\nu]}{2}\right\}.
\end{equation}  
Although all the symmetries in $\Gamma^{(-)}$ are broken as
with the usual Wilson, but the remaining group is larger. 
This symmetry group is interpreted as the subgroup of eight-flavor symmetry 
in the branches.
What we showed here are just examples. The notion of flavored-mass gives us a wide class of
Wilson cousins with desirable numbers of flavors. It is obvious that from the flavored-mass we can also obtain generalized domain-wall and overlap fermions.

{\bf Brillouin fermion :}
In Ref.\cite{DK}, a way of improving dispersion relation in Wilson fermion
was discussed independently. This is called ``Brillouin fermion".
Here the Wilson term is replaced by the other form of the covariant laplacian,
which minimizes the breaking of the rotational symmetry near the Brillouin boundary.
From the viewpoint of the flavored-mass terms, the Brillouin fermion corresponds to
the case of the total sum of the four flavored-mass terms as
\begin{equation}
M_{\mathrm V}+M_{\mathrm T} + M_{\mathrm A} + M_{\mathrm P},
\end{equation}
which split the 16 species into 1- and 15-flavor branches.
(We note that a similar form was referred to in the context of improvement of overlap \cite{hi})
We can utilize the 1-flavor branch as Wilson fermion or overlap kernel.
\cite{DK} argues that the merits of this fermion is that not only the dispersion relation 
becomes more continuum-like, but the Dirac spectrum becomes more Ginsparg-Wilson-like.
The calculation in the paper also shows that this setup is more effective as the quark mass is heavier.
It is an interesting course to consider its application to the heavy quark systems.
One concern about this formulation is that the four flavored-mass terms may be renormalized
independently, and the (1,15) splitting could be broken.
The calculation in \cite{DK} shows that it is not the case, and the species-splitting is kept also in 
the interacting theory.

\subsection{Staggered-Wilson fermion} 
\label{SW}

There are also non-trivial flavored-mass terms for staggered fermions, 
which satisfy $\gamma_{5}$ hermiticity and work as a covariant laplacian. 
The staggered fermions with flavored-mass terms was referred to in the classical 
paper by Golterman and Smit \cite{GS}, and recently revisited 
by Adams in \cite{Adams} as extension of Wilson and overlap fermions. 
This setup has been intensively 
studied in several works \cite{Hoel, PdF, CKM2, MNKO, Va}.
One advantage of this formulation is reduction of matrix sizes in the quark propagator, 
which would lead to reduction of numerical costs in lattice QCD with overlap fermions. 
In this subsection we discuss details of progress and problems in this course.

For staggered fermions we only have two 
possibilities of flavored-mass terms satisfying 
$\epsilon\sim\gamma_{5}\otimes\xi_{5}$-hermiticity: 
$\id\otimes\xi_{5}$ and $\id\otimes \sigma_{\mu\nu}$.  
The flavored-mass term proposed by Adams has a former form,
\begin{equation}  
M_{\mathcal A} = \epsilon\sum_{sym} \eta_{1}\eta_{2}\eta_{3}\eta_{4}
C_{1}C_{2}C_{3}C_{4}
= (\id \otimes \xi_{5}) + O(a),
\label{AdamsM}
\end{equation}
with $(\eta_{\mu})_{xy}=(-1)^{x_{1}+...+x_{\mu-1}}\delta_{x,y}$ and
$(\epsilon)_{xy}=(-1)^{x_{1}+...+x_{4}}\delta_{x,y}$.
where the factor $1/24$ is hidden in the symmetric sum $\sum_{sym.}$. 
The four tastes fall into $\xi_{5}=+1$ subspace with two tastes
and $\xi_{5}=-1$ subspace with the other two tastes. 
Indeed, as shown in \cite{Adams, PdF}, 
the degenerate Dirac spectrum is split into two branches.
By introducing proper mass shift, the staggered version of
Wilson fermion, which we call ``staggered-Wilson fermion",
is written as
\begin{equation}  
S_{\mathcal A}\,=\, \sum_{xy}\bar{\chi}_{x}(D_{\mathcal A})_{xy}\chi_{y}\,=\,\sum_{xy}\bar{\chi}_{x}[\eta_{\mu}D_{\mu}
+r(1+M_{\mathcal A})+m_{0}]_{xy}\chi_{y},
\label{AdamsS1}
\end{equation}
We re-derive (\ref{AdamsM}) from the flavored-mass terms for naive fermions.
As shown in \cite{CKM1}, spin diagonalization decomposes $M_{\rm P}$ into 
four equivalent staggered flavored masses as
\begin{equation}
\bar{\psi}_{x}C_{1}C_{2}C_{3}C_{4}\psi_{x}\,\,\,\,\,\,\,\,\,\,\to\,\,\,\,\,\,\,\,\,\,
\pm\bar{\chi}_{x}(\epsilon\eta_{1}\eta_{2}\eta_{3}\eta_{4}C_{1}C_{2}C_{3}C_{4})\chi_{x}.
\label{SD}
\end{equation} 
Indeed, the Dirac spectrum of the staggered-Wilson operator shown in \cite{Adams, PdF}
is the same as the left figure of Fig.~\ref{Mp} up to the degeneracy of the eigevalues.

We now investigate symmetry of the staggered-Wilson fermion. 
We begin with review of the original staggered symmetry \cite{GS},
which is given by
\begin{equation}
\{ C_{0}, \,\Xi_{\mu},\, I_{s},\, R_{\mu\nu} \}\,\,\times\,\,\{U^{\epsilon}(1)\}_{m=0},
\label{sSym}
\end{equation}
where $C_{0}$ is lattice charge conjugation, 
$\Xi_{\mu}$ is is shift transformation, 
$I_{s}$ is spatial inversion, 
$R_{\mu\nu}$ is hypercubic rotation, 
and $U^{\epsilon}(1)$ is residual chiral symmetry: $\chi_{x}\,\to\, e^{\theta\epsilon_{x}}\chi_{x}$.
As well-known, the flavor and rotation symmetries are summarized 
as $\Gamma_{4} \rtimes SW_{4, {\rm diag}}$.
$\Gamma_{4}$ is a Clifford group operating as flavor reflection and
$SW_{4, {\rm diag}}$ is the diagonal hypercubic subgroup of
euclidian rotation $SO(4)$ and flavor $SU(4)$.
By looking into its timeslice group \cite{GS}, it is shown that
15 pseudoscalars of flavor-$SU(4)$ fall into 7 irreducible representations (irreps) 
as ${\bf 1}:\xi_{4},\xi_{45},\xi_{5}, \,\,\,{\bf 3}:  \xi_{i},\xi_{i5},\xi_{ij},\xi_{i4}$.
Now let us move to the symmetry of the staggered-Wilson, 
which was intensively studied in \cite{steve, MNKO}.
The flavored-mass (\ref{AdamsM}) breaks the staggered symmetry down to
\begin{equation}
\{ C_{0}, \Xi_{\mu}', R_{\mu\nu} \},
\label{A1Sym}
\end{equation}
where $\Xi_{\mu}'\,\equiv\, \Xi_{\mu}I_{\mu}$. 
We take a brief look at this symmetry in terms of physical parity,
charge conjugation and the hypercubic symmetry.  (See \cite{steve, MNKO} for details.)
Regarding the parity, the action is invariant under the transformation 
$\Xi_{4}I_{s}\sim(\gamma_{4}\otimes 1)$, which essentially stands for parity transformation.
For charge conjugation, the staggered charge conjugation symmetry 
$C_{0}$ remains intact in this case. Thus, physical charge conjugation for 
the two-flavor branch can be formed in a similar way to usual staggered
fermions as shown in \cite{GS}.
For Lorentz symmetry, the action is invariant under the staggered rotation $R_{\mu\nu}$ 
and the shifted-axis reversal $\Xi_{\mu}'$. These two groups are enough to 
form the proper hypercubic group $SW_{4}$ as with the staggered fermion. 
The staggered-Wilson action thus possesses charge conjugation,
parity and hypercubic symmetry.
This result means that the staggered-Wilson satisfies the minimum conditions 
for practical use in lattice QCD.


We now need to figure out advantages of staggered-Wilson and staggered-overlap over the originals. 
The staggered-Wilson fermion has been studied from several viewpoints,
including the index theorem, Aoki phase, numerical costs and taste-breaking. 

{\bf Index theorem and overlap kernel :}
The study on the index theorem in the staggered-Wilson fermion via the spectral flow of the hernitian Dirac operator, which means the net number of near-origin eigenvalue crossings counted with signs of slopes, was first done by Adams in \cite{Adams}, and followed by \cite{CKM1, PdF, Va}.
In this case the index of the Dirac operator is defined as 
${\rm Index}(D_{\mathcal A}) = -{\rm Spectral\,\,\,flow}(H_{\mathcal A})$ 
with $H_{\mathcal A}(m_{0})=\epsilon D_{\mathcal A}(m_{0})$, and it was shown
that the index correctly detects the gauge topology in this setup.
The stability of separation of low and high crossings in eigenvalue flow 
was also discussed, which indicates
applicability of $H_{\mathcal A}(m_{0}<0)$ as overlap kernel.
In \cite{CKM1} the index theorem for the generalized Wilson including
the Brillouin fermion was also studied.

{\bf Aoki phase :}
The parity phase structure for the staggered-Wilson fermion was
studied by using the Gross-Neveu model in \cite{CKM2} and strong-coupling
lattice QCD \cite{MNKO}.
The GN study shows that the parity-broken phase and the 2nd-order critical line
exist as with the case of the usual Wilson fermion,
which implies possible Aoki scenario in this formulation.
The strong-coupling lattice QCD also indicates the existence of the massless
pion and the PCAC relation near the parity phase boundary.
We expect the lattice chiral perturbation works to study this topic further.

{\bf Numerical costs :}
One of possible advantages of the staggered-Wilson fermion 
is that it could reduce the computation expense of overlap fermions.
The computational costs in the staggered-overlap quark propagator 
were well studied in \cite{PdF}. According to this work, there is competition 
between an advantage from the 4-times smaller matrix to invert and a disadvantage 
from 4-link hopping terms in the flavored mass.
Since the smaller matrix size requires fewer matrix-vector multiplications 
in calculation of the sign function,
the total CPU cost becomes smaller in the staggered-overlap at least for the free theory. 
However, this merit is ruined by large gauge fluctuation
due to the 4-link transporters. 
The gauge fluctuation reduces the splitting of the two branches, 
and unitary projection and inversion get more difficult.
As a result, the reduction of the total CPU time is about $\mathcal{O}(2)$ for $\beta=6$.
Thus, so far, the numerical advantage of the staggered-overlap over the usual overlap
is not so significant. We however note that we can improve staggered-Wilson setups in terms of both
staggered \cite{Va} and Wilson. The numerical cost would be reduced through the improved versions of the staggered-Wilson.

 {\bf Taste breaking :}
It is obvious that the two flavors in staggered-Wilson suffer $SU(2)$ taste symmetry breaking. 
But it does not necessarily mean the mass-splitting of the associated three pions $\pi_{0},\pi_{\pm}$. 
The pion mass splitting depends on the remaining discrete flavor symmetry:
If this symmetry is large enough, 
we have degenerate three pions even at the finite lattice spacing.
Recently Ref.~\cite{steve} has reported that the classification of operators 
by the timeslice symmetry indicates the three-degenerate pions: 
For pseudoscalar mesonic operators in staggered-Wilson, 
the 7 irreps and flavor-singlet operators of the original staggered are mixed in $\xi_{5}$ pairs.
And by focusing on the physical two-flavor branch $\ell$ we find the three pion states 
$\pi_{0}, \pi_{\pm}$ are in the 3d irreps as $\bar{\ell} (\gamma_{5}\otimes\sigma_{i})\ell\,\,(i=1,2,3)$. 
The same work also shows that, in the pion potential of the stggered-Wilson chiral Lagrangian, 
the taste breaking starts from $O(a^{4})$ and $O(a^{2}m)$.
This result implis that the two flavors in the staggered-Wilson fermion 
suffer relatively small flavor breaking and suits the two-flavor lattice QCD.

{\bf Another type:}
In \cite{Hoel, PdF}, the other type of staggered-Wilson fermions were proposed.
This type corresponds to the case with $\id\otimes \sigma_{\mu\nu}$, the
only other possibility satisfying the staggered $\gamma_{5}$-hermiticity.
This flavored mass is given by
\begin{equation}  
M_{\mathcal{H}}=i(\eta_{12}C_{12}+\eta_{34}C_{34})
= [\id \otimes (\sigma_{12}+\sigma_{34}) ] + \mathcal{O}(a)\, ,
\label{HoelM}
\end{equation}
with
$(\eta_{\mu\nu})_{xy}=\epsilon_{\mu\nu}\eta_{\mu}\eta_{\nu}\delta_{x,y}$,
$(\epsilon_{\mu\nu})_{xy}=(-1)^{x_{\mu}+x_{\nu}}\delta_{x,y}$,
$C_{\mu\nu}=(C_{\mu}C_{\nu}+C_{\nu}C_{\mu})/2$.
We note that the original flavored mass proposed by Hoelbling in \cite{Hoel} has a different form,
but we here discuss the above form since it has larger discrete symmetry \cite{CKM1, steve}.
This flavored mass splits four tastes into three branches: one with positive mass, two with zero mass
and the other one with negative mass. 
It thus produces single-flavor staggered-Wilson fermion, which seems to have wider applicability.
However, the symmetry of the staggered fermion with this flavored mass term is small as
$\{ C_{T}, \Xi_{\mu}', R_{12}, R_{34}, R_{24}R_{31} \}$
with $C_{T}\equiv R_{21}R_{13}^{2}C_{0}$.
Breaking of rotation symmetry indicates that we need to tune parameters to restore Lorentz symmetry.
Indeed the recent study on symmetries of staggered-Wilson fermions by Sharpe \cite{steve}
reports that recovery of Lorentz symmetry requires fine-tuning of parameters 
in the gluonic sector in lattice QCD.


\section{Central Branch}
\label{sec:CB}

In this section we discuss another way of use of flavored-mass or Wilson fermions.
We usually concentrate on the one-flavor edge branch of the Wilson Dirac spectrum. 
However, the flavored-mass fermions without onsite terms 
(or species-singlet mass term) have a larger symmetry in general.
In terms of the flavor-chiral symmetry of naive fermions in (\ref{sym_naive_all}), 
the onsite term ($\sim\bar{\psi}_{n}\psi_{n}$) breaks the invariance under 
any transformation of the axial-type group $\Gamma^{(-)}_{\A}$ in Eq.(\ref{sym_t-}).
Thus, dropping onsite terms can restore some invariance under the group,
and the action comes to have a larger symmetry, including axial symmetry.

\begin{figure}
\begin{minipage}{0.33\hsize}
\begin{center}
\includegraphics[bb=0 100 1024 700, clip, width=4cm]{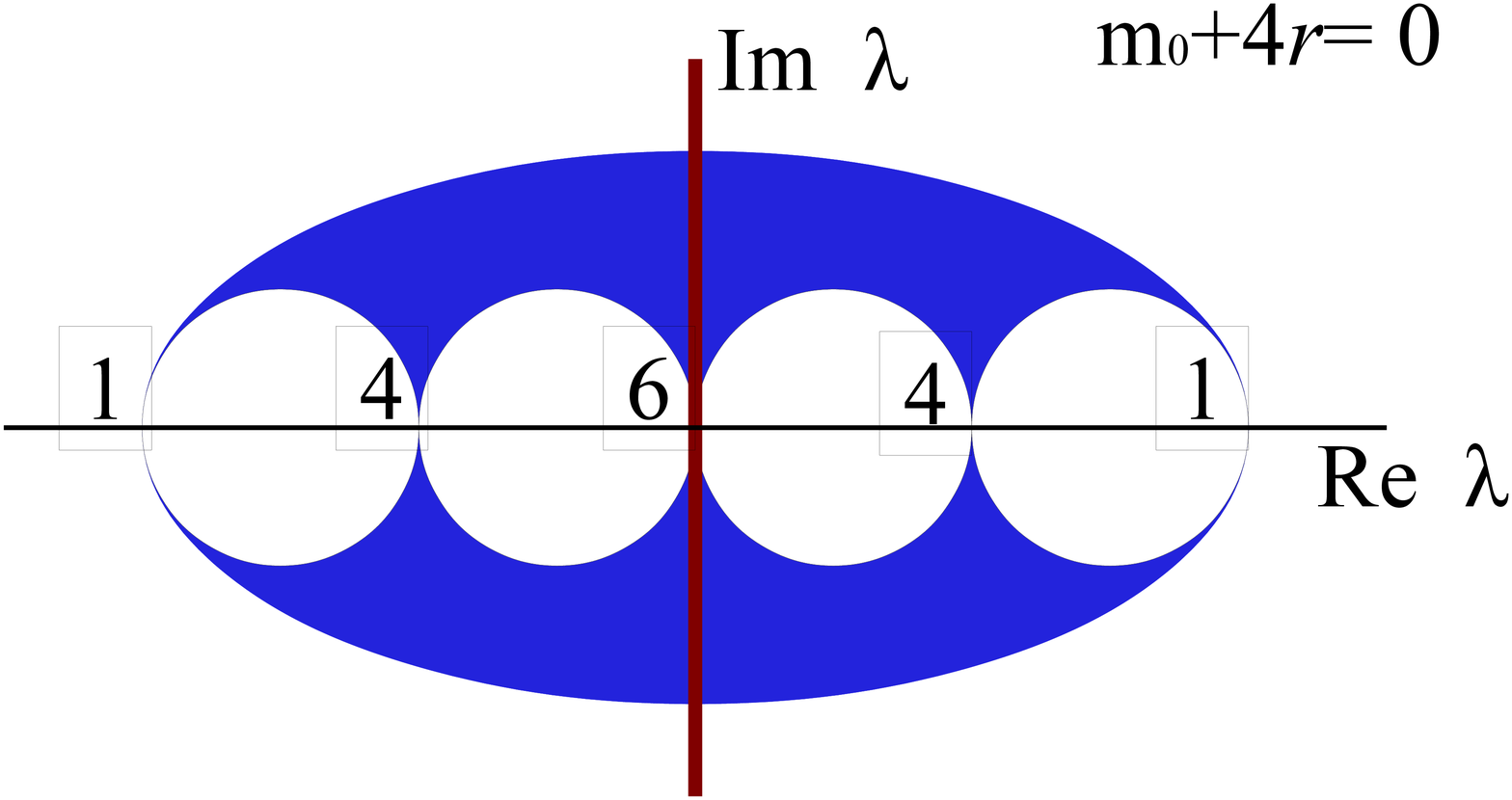} 
\end{center}
\caption{The Wilson Dirac spectrum with $m_{0}+4r=0$. 
The central branch crosses the origin. }
\label{WilD}
\end{minipage}
\begin{minipage}{0.33\hsize}
\begin{center}
\includegraphics[bb=0 0 1024 500, clip, width=5.5cm]{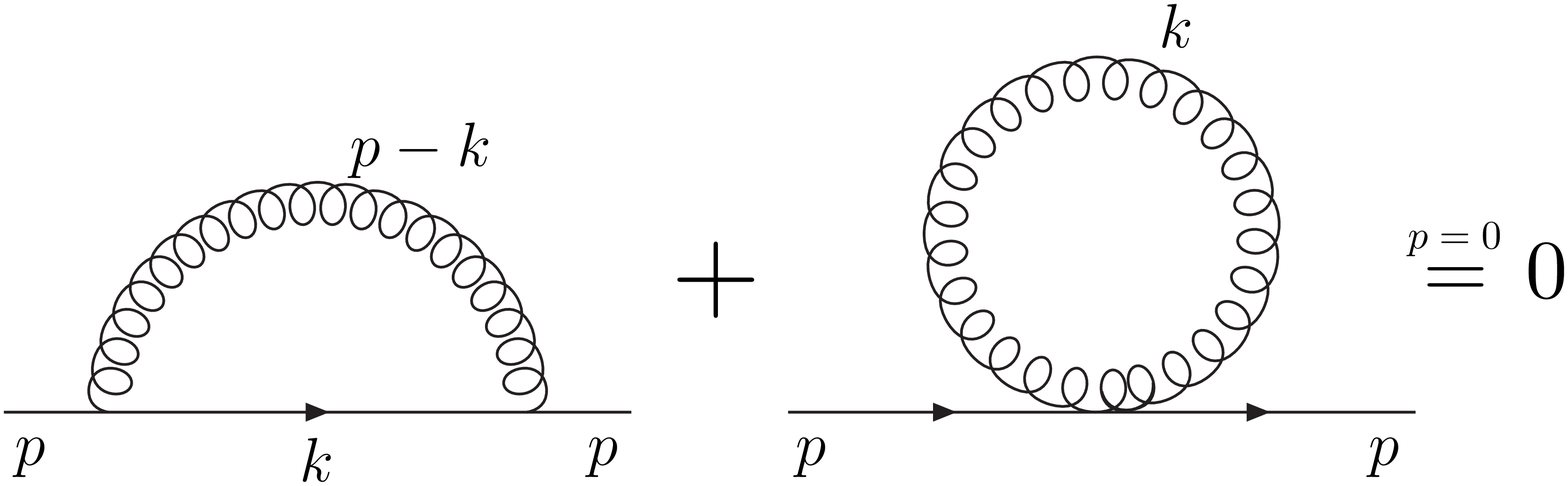} 
\end{center}
\caption{The diagrams contributing to the 1-loop fermion self-energy. The additive mass renormalization from these diagrams becomes zero. }
\label{diag}
\end{minipage}
\begin{minipage}{0.33\hsize}
\begin{center}
\includegraphics[width=4.5cm]{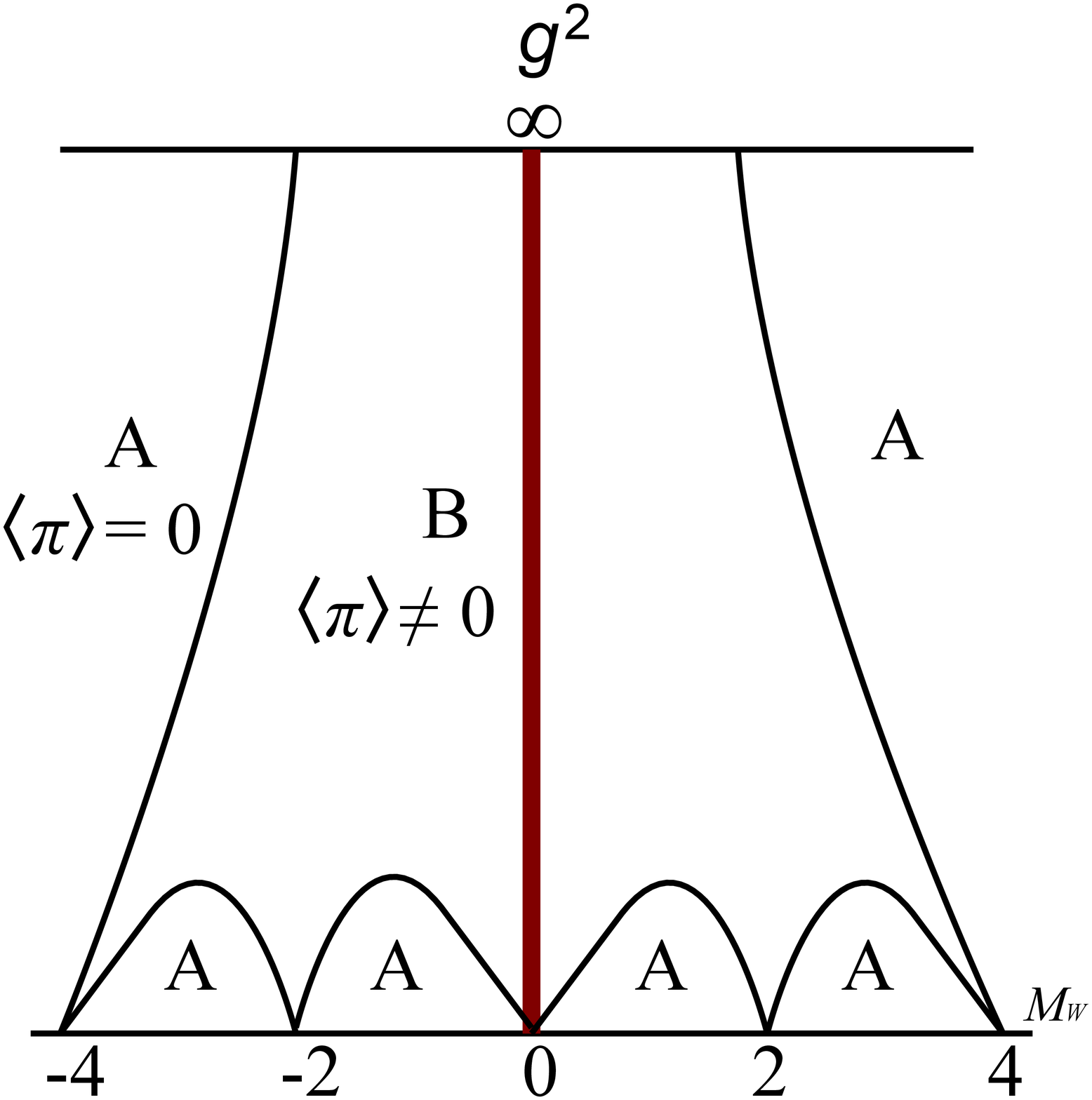} 
\end{center}
\caption{The Aoki phase diagram with Wilson fermions. The red line corresponds to
the central branch, where the extra symmetry emerges.}
\label{WilA}
\end{minipage}
\end{figure}

In \cite{CKM2} and \cite{Rev}, it was shown that the Wilson fermion with the 
condition $m_{0}+4r=0$ has an extra $U(1)$ symmetry besides the usual $U(1)$ baryon symmetry.
The condition corresponds to the six-flavor central branch of the Wilson Dirac spectrum 
as shown in Fig.~\ref{WilD}. We call it ``central-branch fermion".
The action in this case is given by
\begin{equation}
S_{\rm CB}=\sum_{n,\mu}\bar{\psi}_{n}\gamma_{\mu}D_{\mu}\psi_{n}-r\sum_{n,\mu}\bar{\psi}_{n}C_{\mu}\psi_{n}.
\label{CB}
\end{equation}
In this case the $U(4)\times U(4)$ symmetry of naive fermions in (\ref{sym_t+})(\ref{sym_t-}) 
are broken down to
\begin{equation}
\Gamma^{(+)}\,\,\to\,\,\left\{\mathbf{1}_4 \right\},\,\,\,\,\,\,\,\,\,\,
\Gamma^{(-)}\,\,\to\,\,\left\{(-1)^{n_1+\ldots+ n_4}\right\}.
\label{CBsym}
\end{equation}  
The usual Wilson fermion has only the vector symmetry under $\mathbf{1}_4 \in\Gamma^{(+)}$.
The invariance under $(-1)^{n_1+\ldots+ n_4}\in\Gamma^{(-)}$
is restored only with the central-branch condition $m_{0}+4r=0$.
It is notable that the form $(-1)^{n_1+\ldots+ n_4}$ is the same as
the staggered chiral transformation $U^{\epsilon}(1)$, which works as
the axial rotation as shown in (\ref{sym_naive_all}).
It is obvious that the quark mass term $\bar{\psi}\psi$ is not invariant under this,
and the extra symmetry prevents the quark mass term $\bar{\psi}\psi$
from being generated via loop effects.
It means that the additive mass renormalization cannot occur in this case.
Although the central branch of the usual Wilson fermion has six flavors,
the symmetry enhancement on the central branch is generic with the flavored-mass fermions
and this fact inspires us to consider new realizations of chiral fermions.

The one-loop lattice perturbation in \cite{MKL} shows that
the additive quark mass renormalization becomes zero unlike the usual Wilson. 
The quark self-energy at one-loop level in a massless case is in general written as
\begin{equation}
{g_{0}^2\over{16\pi^{2}}}\left( {\Sigma_{0}\over{a}} + i\gamma_{\mu}p_{\mu}\Sigma_{1}\right).
\end{equation}
We now focus on $\Sigma_{0}$, which is composed of the sunset $\Sigma_{0}^{(\alpha)}({\rm sun})$ and tadpole $\Sigma_{0}^{\alpha}({\rm tad})$ diagrams.
$\alpha=1,2,...,6$ identifies the six poles of the propagator, which we denote as 
$\pi_{\mu}^{(1)}=(0,0,\pi,\pi)$, $\pi_{\mu}^{(2)}=(0,\pi,0,\pi),\cdot\cdot\cdot$. 
For the central-branch fermion (\ref{CB}),
$\Sigma_{0}^{(\alpha)}({\rm sun})$ is calculated as
\begin{align}
\Sigma_{0}^{(\alpha)}({\rm sun})&= {g_{0}^{2}\over{a}}C_{F}\int^{\pi}_{-\pi}{d^{4}k\over{(2\pi)^{4}}}\sum_{\rho} { (\sin^{2}{k_{\rho}\over{2}}-\cos^{2}{k_{\rho}\over{2}})(-\sum _{\lambda}\cos k_{\lambda}) 
+\sin^{2} k_{\rho}  
\over{4(\sum_{\lambda}\sin^{2} {k_{\lambda}\over{2}})(\sum_{\mu}\sin^{2}k_{\mu}+(-\sum _{\mu}\cos k_{\mu})^{2})}}e^{i\pi_{\rho}^{(\alpha)}}\,\,\,\,=\,\,\,0,
\end{align} 
where $e^{i\pi_{\rho}^{(\alpha)}}$ takes $\pm 1$ depending on $0$ or $\pi$ locations of
the poles in the direction $\rho$. This sign difference leads to cancelation between 
dimensions $\rho$ in the integral for any pole $\alpha$.
The total contribution from the tadpole diagram, which we denote $I^{(\alpha)}({\rm tad})$, is given by 
\begin{equation}
I^{(\alpha)}({\rm tad}) =\int^{\pi}_{-\pi}{d^{4}k\over{(2\pi)^{4}}}
{-g_{0}^{2}\sum_{a}\{T^{a},T^{a}\}_{cc}  \over{8a\sum_{\lambda}\sin^{2}k_{\lambda}/2}}
\sum_{\rho}\left(-i\gamma_{\rho}\sin (ap_{\rho}+\pi^{(\alpha)}_{\rho})+\cos (ap_{\rho}+\pi^{(\alpha)}_{\rho})\right)\,\,\,\propto\,\,\, i\gamma_{\mu}p_{\mu}
\end{equation}
where we again have cancelation between dimensions $\rho$ in $\Sigma_{0}^{(\alpha)}({\rm tad})$,
which means $\Sigma_{0}^{(\alpha)}({\rm tad})=0$ for any of the six poles. 
In the end the total $\Sigma_{0}$ becomes zero for each of the six Dirac poles as
$\Sigma_{0}^{(\alpha)}=\Sigma_{0}^{(\alpha)}({\rm sun})+\Sigma_{0}^{(\alpha)}({\rm tad})=0$.
This verifies absense of additive renormalization in the setup.

One concern on this setup is how the parity-broken phase affects the central branch.
The central branch ($m_{0}+4r=0$) corresponds to the central cusp 
in the parity phase diagram in Fig.~\ref{WilA}.
It is believed that this point is located within the parity broken phase
at strong and middle gauge couplings, 
and the six flavors may be mutilated.
Not to mention, we do not need to consider this problem if we take the chiral limit from 
the parity-symmetric phase to the central branch as shown in \cite{PdF}.    
Nevertheless, we will find an important aspect of the central branch fermion by looking into
the relation with the Aoki phase.
The strong-coupling and large-$N_{c}$ limit of lattice QCD right on the Wilson
central branch was studied in \cite{Rev}.
It shows that the condensates in the strong-coupling limit of the central branch 
are given by 
\begin{equation}
\sigma = \dfrac{M_W}{4r^2} \,,\,\,\,\,\,\,\,\,\,\,\,\,\,  \pi =\dfrac{1}{16r^4(1+r^2)} (8r^4 - M_W^2(1+r^2)), 
\end{equation}
where $\sigma$ and $\pi$ essentially stand for chiral $\langle\bar{\psi}\psi\rangle$ and pion $\langle\bar{\psi}\gamma_{5}\psi\rangle$ 
condensates with $M_{W}=m_{0}+4r$. It also shows that one of mesonic excitations has mass 
in the form as
\begin{eqnarray}
\cosh (m_{SPA}\,) = 1 + \frac{2 M_W^2(16+M_W^2)}{16-15M_W^2}.
\label{dispersion}
\end{eqnarray} 
For the central branch $M_{W}=m_{0}+4r=0$, we have $\sigma=0, \pi=1/2(1+r^2)$ and $m_{SPA}=0$.
This result means that the extra symmetry is spontaneously broken
and it leads to a massless NG meson, which seems to mimic QCD successfully. 
However, the symmetry is not broken by the chiral condensate $\langle \bar{\psi}\psi\rangle$
but by the parity broken condensate $\langle \bar{\psi}\gamma_{5}\psi\rangle$.
Rather, the chiral condensate becomes exactly zero on the central branch.
To summarize, Strong-coupling lattice QCD on the central-branch fermion has the condensate as
$\langle \bar{\psi}\psi\rangle=0,\,\langle \bar{\psi}\gamma_{5}\psi\rangle\not=0$.
This is a situation as if the roles of operators $\bar{\psi}\psi$ and 
$\bar{\psi}\gamma_{5}\psi$ are exchanged.
In other words, the mass basis in this case is different from the usual QCD
and we need to introduce a quark mass term in a twisted form, e.g. $m\bar{\psi}\gamma_{5}\psi$.
This fact indicates that the central-branch fermion is deeply related to
the twisted-mass Wilson fermion. 
We note that the maximally-twisted-mass Wilson fermion is regarded as the average of
the two one-flavor edge branches, and the central branch is 
located at the center point between these two.
It implies that the central branch fermion can be seen as an automatic realization of 
the maximally-twisted-mass Wilson fermion without $O(a)$ discretization errors.
If this view is correct, we can expect the parity breaking disappears in the continuum limit 
as in the twisted-mass Wilson.


We have another concern on this case : positivity of determinant. In the case of the maximally-twisted-mass Wilson QCD, the traceless flavor matrix $\tau_{3}$ in the twisted-mass term $i\bar{\psi}\gamma_{5}\tau_{3}\psi$ works to keep the quark determinant positive and avoid the theta term. Unlike this case, the quark condensate $\langle\bar{\psi}\gamma_{5}\psi\rangle$ in the strong-coupling limit of the central branch has no traceless structure in the 6-species flavor space. This problem is deeply related to the question how the $U(1)$ problem is incorporated in this formulation. However, we note the chiral symmetry in this setup is related to the generator $(-1)^{n_1+\ldots+ n_4}$ instead of naive $\gamma_{5}$. Thus, if the true condensate at the weak coupling has a form $\bar{\psi}(-1)^{n_1+\ldots+ n_4}\psi$ and it has traceless flavor structure for the six flavors, we may not necessarily have the issue. We have no consensus on this problem yet, and we hope detailed numerical study can answer this question. In any case, we can bypass the issue by introducing the pair of the central-branch fermions with $\tau_{3}$ introduced in the mass term. In this case, the setup contains 12 flavors. This can be an alternative way for studying 12-flavor $SU(3)$ gauge theory, whose existence of the infrared fixed point is now controversial. Not to mention, we are free from the problem when we take the chiral limit from the parity-symmetric phase as we discussed.

So far we have concentrated on the central branch in the usual Wilson fermion.
However, the flavored-mass fermions discussed in the previous section
in general enjoy the extra symmetry restoration when the onsite term is dropped.
For example, the fermion action with $M_{\rm P}$ but without mass shift
has larger symmetry than (\ref{sym_t+P}):
the flavor-chiral symmetry of $\bar{\psi}_{n}(\gamma_{\mu}D_{\mu}+M_{\rm P})\psi_{n}$ is
\begin{equation}
\Gamma^{(+)}\to \left\{\mathbf{1}_4\,,\,\, (-1)^{n_1+\ldots+ n_4}\gamma_5
\,,\,\,(-1)^{n_{\mu,\nu}}\frac{i \,[\gamma_\mu\,,\gamma_\nu]}{2}\right\},\,\,\,\,\,\,\,\,\,\,\,\,\,\,\,
\Gamma^{(-)}\to \left\{(-1)^{n_\mu}\gamma_\mu\,,\,\,(-1)^{\check{n}_\mu}i \gamma_\mu\gamma_5
\right\}.
\end{equation}  
Restoration of the invariance under the axial-type group $\Gamma^{(-)}$
indicates absence of additive mass renormalization again.
However, in this case, we have no central branch as seen in the left figure of Fig.~\ref{Mp} 
and we cannot take advantage of the larger symmetry.
The staggered-Wilson fermion in (\ref{AdamsS1}) also has the similar property:
$S_{\mathcal A}$ in (\ref{AdamsS1}) without mass shift
has larger symmetry as
\begin{equation}
\{ C_{0}, \,\,C_{T}'\Xi_{\mu},\,\, C_{T}'I_{s},\,\, R_{\mu\nu} \}.
\label{A0Sym}
\end{equation}
where $C_{T}'$ is given as a special charge conjugation 
$C_{T}':\,\,\chi_{x}\,\to\,\bar{\chi}_{x}^{T},\,\,
\bar{\chi}_{x}\,\to\,\chi_{x}^{T},\,\,
U_{\mu,x} \,\to\, U_{\mu,x}^{*}$.
Again, this case is not useful since there is no central branch. 
However the situation is different for the other type of 
the staggered-Wilson fermion in (\ref{HoelM}) \cite{PdF,CKM2}:
the symmetry of the action (\ref{HoelM}) without mass shift becomes larger \cite{steve} as
\begin{equation}
\{ C_{T}, \,\,C_{T}',\,\, \Xi_{\mu}', \,\,R_{12},\,\, R_{34},\,\, R_{24}R_{31} \}.
\end{equation}
The point is that we have a two-flavor central branch in this case. 
The extra symmetry is special charge conjugation $C_{T}'$,
which prohibits the additive mass renormalization.
The numerical calculation for this case was performed in \cite{PdF}.
Although this fermion has possibility of being a two-flavor setup without
necessity of mass parameter fine-tuning,
we have necessity of other parameter tunings for Lorentz symmetry for this case
as we discussed \cite{steve}, which compensates the advantage.
For our reference, the discrete symmetry of the staggered kinetic term, the
two flavored-mass terms and the usual mass term are summarized in Table.\ref{symT}.
\begin{table}[tbp]
\begin{center}
{\tabcolsep = 1 em
\scalebox{0.8}{
\begin{tabular}{ccccccc} \hline\hline
 & $C_{T}'$ &  $\Xi_{\mu}$ & $I_{\mu}$ & $C_{T}'\Xi_{\mu}$ & 
$C_{T}'I_{\mu}$ & $\Xi_{\mu}I_{\mu}$ \\ \hline
$S_{st}$ & $\circ$ & $\circ$ & $\circ$ & $\circ$ & $\circ$ & $\circ$ \\
$\bar{\chi}_{x}M_{\mathcal A}\chi_{x}$ & $\times$ & $\times$ & $\times$ & $\circ$ & $\circ$ & $\circ$ \\
$\bar{\chi}_{x}M_{\mathcal H}\chi_{x}$ & $\circ$ & $\times$ & $\times$ & $\times$ & $\times$ & $\circ$ \\
$\bar{\chi}_{x}\chi_{x}$ & $\times$ & $\circ$ & $\circ$ & $\times$ & $\times$ & $\circ$ \\ \hline\hline
\end{tabular}
}
}
\end{center}
\caption{}
\label{symT}
\end{table}

The interesting goal in this direction is to construct a two-flavor central-branch 
fermion by using a proper combination of the four flavored-mass terms 
shown in (\ref{F-mass}).
In principle, it is possible to find a combination of flavored mass 
which has a two-flavor central branch.
However, the question is whether hypercubic symmetry
holds in such a case. If it is broken as with the single-flavor staggered-Wilson 
fermion in (\ref{HoelM}), it is not useful. 
We need further study to answer this question.


\section{Minimal-doubling}
\label{sec:MD}

\begin{figure}
\begin{center}
\includegraphics[bb=0 300 1024 550, clip,width=8cm]{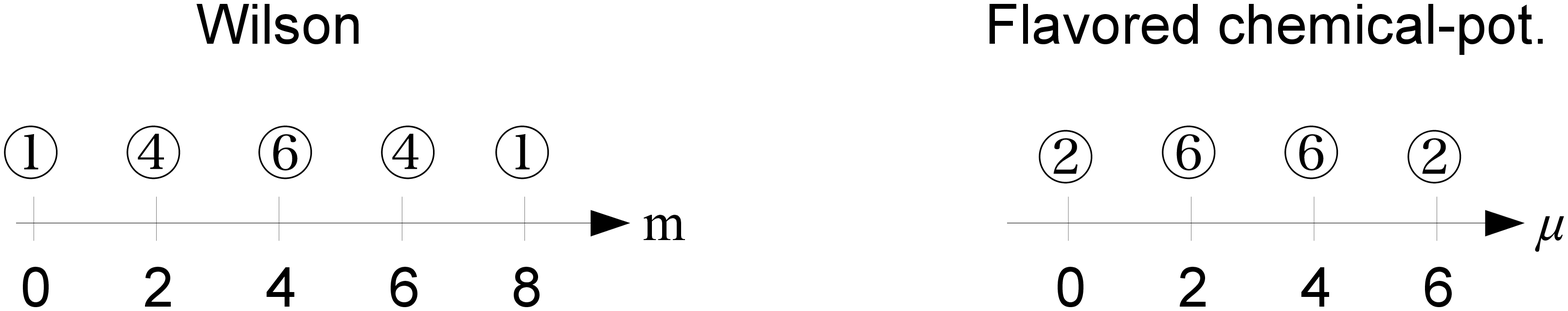} 
\end{center}
\caption{Species-splittings in Wilson and Karsten-Wilczek fermions.
Circled numbers stand for the number of massless flavors on each point.\cite{misumi}}
\label{MD}
\end{figure}

\begin{figure}
\begin{center}
\includegraphics[bb=0 80 1024 670, clip, width=6cm]{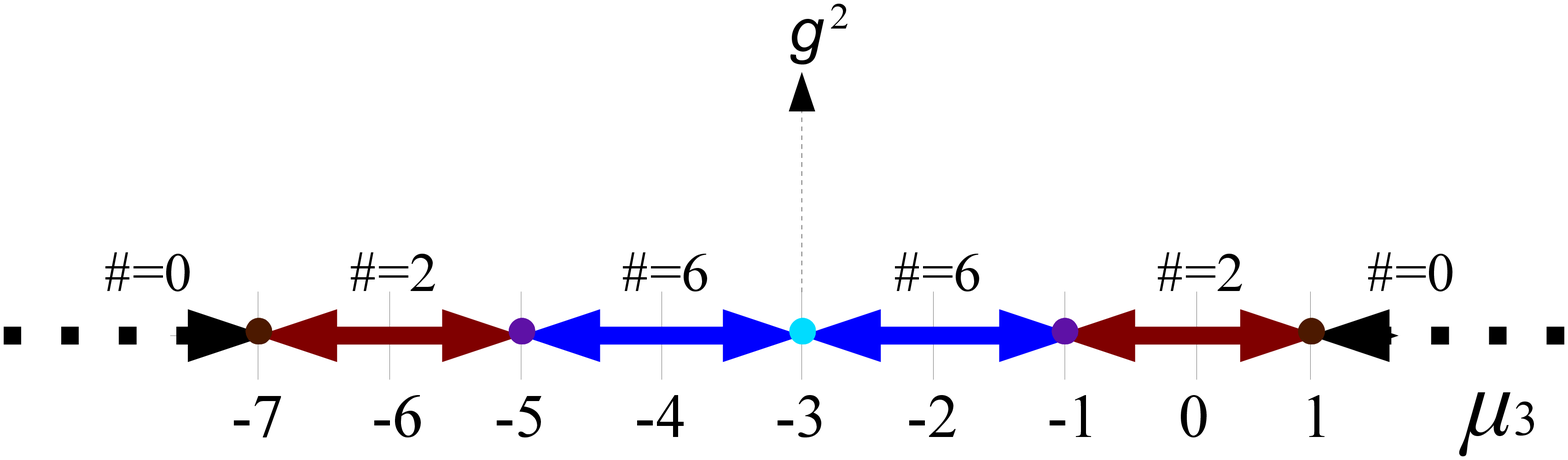} \qquad
\includegraphics[bb=0 80 1024 670, clip, width=6cm]{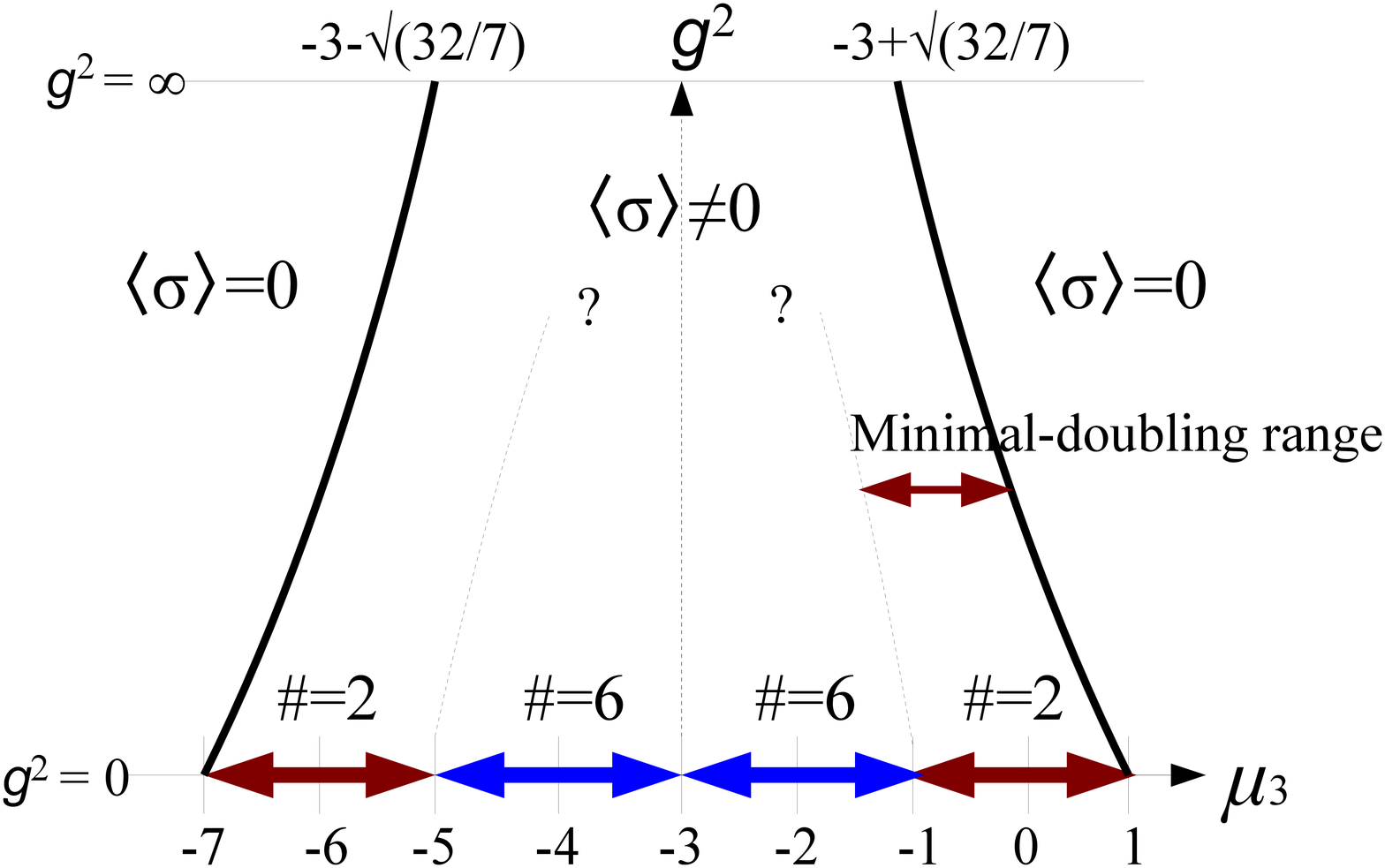} 
\end{center}
\caption{(Left) The number of flavors in free Karsten-Wilczek fermion as 
a function of the parameter $\mu_{3}$. See \cite{misumi} for details. (Right) Conjecture on $\mu_{3}$-$g^{2}$ chiral phase diagram for Karsten-Wilczek fermion with r=1 and $d_{4}=0$ \cite{misumi}. }
\label{p-range}
\end{figure}

In this section we discuss another formulation with the desirable number of flavors and chiral symmetry.
One drawback in the flavored-mass fermions in Sec.~\ref{sec:FM} and \ref{sec:CB} 
is that it explicitly breaks all the axial symmetry except on the central branches.
The interesting question is whether we can reduce species to two with keeping the remnant 
of chiral symmetry.
We now consider free Dirac operators with the flavored mass $M_{F}$ multiplied 
by $i\gamma_{4}$,
\begin{equation}
aD_{\rm fc}(p)=i\gamma_{\mu}\sin p_{\mu}a\,+i\gamma_{4}M_{F}(p),
\label{FCP}
\end{equation}
where we assume $M_{F}$ has a hypercubic or cubic symmetric form.
Here the degeneracy of 16 species is not lifted by the flavored-mass term, 
but by specie-dependent imaginary chemical potential terms which we call
 ``flavored-chemical-potential(FCP)" \cite{misumi}.
Although we can consider a real type of FCP terms,
we concentrate on the imaginary type to avoid the sign problem.
(See \cite{misumi, MKO} for real FCP.)
It is notable that this Dirac operator has ultra-locality and exact chiral symmetry,
\begin{equation}
\{\gamma_{5},D_{\rm fc}(p)\}=0.
\end{equation}
In this setup it is possible to reduce 16 species to two without losing all chiral symmetries, 
which is the minimal number allowed by the no-go theorem.
On the other hand, the chemical potential term breaks the lattice discrete symmetry to the subgroup as
cubic symmetry, CT and P \cite{Bed}, which give rise to necessity of fine-tuning several parameters for a correct continuum limit \cite{Cap}. In other words, instead of chiral symmetry breaking in Wilson fermion leading to the mass parameter tuning, FCP fermions have breaking of discrete symmetry leading to necessity of other parameter tuning.
To look into details of this setup, we introduce one concrete example
\begin{eqnarray}
 S_{\mathrm{KW}} & = &
  \sum_{n} 
  \Bigg[
   \bar\psi_n \gamma_\mu D_{\mu}\psi_{n}+m_{0}\bar{\psi}_{n}\psi_{n} 
  + r\sum_{j=1}^3 \bar \psi_n i\gamma_4 (1-C_{j}) \psi_{n}
     + \mu_{3}\bar{\psi}_{n}i\gamma_{4}\psi_{n}
+  d_{4}\bar\psi_x \gamma_{4}D_{4} \psi_{n}
  \Bigg],
\label{Smd}  
\end{eqnarray}
where we introduce a Wilson-like parameter $r$. 
The dimension-3 and dimension-4 counterterms with parameters $\mu_{3}$ and
$d_{4}$ are also introduced for the later discussion on tuning.  
The associated free and massless Dirac operator in momentum 
space is
\begin{equation}
 aD_{\mathrm{KW}}(p) =
  i \sum_{\mu=1}^4 \gamma_\mu \sin ap_\mu
  + i\gamma_4
(\mu_{3}+3r-  r  \sum_{j=1}^{3}\cos ap_j + d_{4}\sin ap_{4}).
 \label{Smdp} 
\end{equation}
For simplicity we first consider $\mu_{3}=0$, $d_{4}=0$ and $r=1$. 
In this case we have only two zeros of the Dirac operator $p=(0,0,0,0)$ and $p=(0,0,0,\pi/a)$ 
while the other 14 species have $O(1/a)$ imaginary chemical 
potential ($\sim {1\over{a}}\bar{\psi}_{n}i\gamma_{4}\psi_{n}$) 
and are decoupled in the naive continuum limit.
In Fig.~\ref{MD} we compare species-splitting of KW fermions in the chemical potential direction
to that of Wilson fermion in the mass direction.
For general values of $\mu_{3}$, the number of physical flavors 
depends on $\mu_{3}$ as shown in the left 
of Fig.~\ref{p-range}: 0 flavor in $\mu_{3}>1$, 2 flavors in $-1<\mu_{3}<1$, 
6 flavors in $-3<\mu_{3}<-1$, 6 flavors in $-5<\mu_{3}<-3$
2 flavors in $-7<\mu_{3}<-5$, 0 flavor in $\mu_{3}<-7$.
The action (\ref{Smd}) is known as the Karsten-Wilczek (KW) fermion 
\cite{KW}, which is the first type of ``minimally doubled fermions"
\cite{KW, CB, CM}.
Although we concentrate on the KW fermion all through this section,
there are other types including Borici-Creutz 
\cite{CB} and Twisted-ordering \cite{CM} fermions. The properties
shown in this section are qualitatively common with all types.

The symmetry of minimally doubled fermions was intensively studied in Ref.\cite{Bed}.
As we have seen in a generic form of the action (\ref{FCP}), 
the KW fermion has one exact chiral symmetry, ultra-locality, spatial cubic symmetry, CT and P. 
Because of this lower discrete symmetry, we need to fine-tune parameters 
to take a correct continuum limit.
The process of the tuning was well discussed in  \cite{Cap}.
In the case of KW fermion, it is pointed out that we need to tune parameters for a dimension-3 
($\bar{\psi}i\gamma_{4}\psi$) and two dimension-4 ($\bar{\psi}\gamma_{4}\partial_{4}\psi$, 
$F_{j4}F_{j4}$) counterterms. Two of these counterterms are shown in the action (\ref{Smd}), 
and we need one more counterterm for the plaquette action.
The necessity of three-parameter tuning seems to compensate 
the merits of the minimal-doubling fermions (chiral, ultra-local and two-flavor),
and they have not been used widely.

However, as seen from the generic argument below (\ref{FCP}),
minimal-doubling fermions are seen as lattice fermions with species-dependent 
chemical potential terms. It is notable that the discrete symmetry 
of minimal-doubling fermions is the same as 
the usual hypercubic-symmetric lattice fermions with the chemical potential,
e.g. naive fermions with the chemical potential
$S_{\rm N}(\mu) =\sum_{n} 
( \bar\psi_n \gamma_j
D_{j}\psi_{n}+\bar\psi_n \gamma_4
\left(e^{\mu}U_{n,n+4} \psi_{n+4} 
-e^{-\mu}U_{n,n-4}\psi_{n-4}\right)/2)$,
where the discrete symmetry is broken down to spatial cubic symmetry, CT and P again. 
Therefore, from the viewpoint of universality class, it is likely that 
both setups belong to the same universality class.
We expect the symmetry of lattice QCD with minimal-doubling fermions will be enhanced to 
that of the two-flavor in-medium QCD (spatial rotation, P, CT,  $SU(2)$flavor-chiral) 
in a continuum limit. Thus it is natural to regard the minimal-doubling fermion as the finite-density 
system and consider the application to the finite-($T$,$\mu$) QCD.
When an effective way of bypassing the sign problem is found in the near future,
the two-flavor lattice fermion suitable for the finite-density system 
with ultra-locality and chiral symmetry will play an important role.

Nevertheless, the minimal-doubling fermions and naive fermions with chemical potential
have difference in the way of introducing the chemical potential.
As seen from the discussion in the beginning of this section, 
the minimal-doubling fermion decouples 14 doublers with $O(1/a)$ imaginary chemical potential
by introducing the FCP term, which is roughly given by the Wilson term $\bar{\psi}M_{F}\psi$
incorporating $i\gamma_{4}$.
Thus, instead of the additive mass renormalization in Wilson fermion, 
the unphysical $O(1/a)$ chemical potential renormalization occurs 
in the interacting minimal-doubling fermion even if we consider it as the in-medium system.
Then we need to fine-tune the dimension-3 parameter $\mu_{3}$ to realize 
physical $O(1)$ chemical potential as we needed to fine-tune the mass 
parameter $m_{0}$ to realize physical quark mass in Wilson fermion. 
This necessity of $\mu_{3}$ tuning is also understood from the well-known fact that 
the naive introduction of chemical potential on the lattice gives rise to 
the unphysical divergence of energy density and requires a counterterm since the violation of
the abelian gauge invariance takes place \cite{HK}. 

This large renormalization leads to instability of the number of
flavors in the interacting theory since the $\mu_{3}$ dependence of flavor number 
in the left of Fig.~\ref{p-range} in a free theory changes due to the renormalization.
In Ref.~\cite{misumi}, the $\mu_{3}$-$g^{2}$ chiral phase structure for KW fermion is studied by using
the Gross-Neveu model and strong-coupling lattice QCD (the right figure of Fig.~\ref{p-range}).
As seen from this, the $\mu_{3}$ range with physical flavors gets narrower 
as the gauge-coupling gets larger, and the minimal-doubling range also 
moves and gets narrower with nonzero gauge coupling.
It means that we at least need to adjust $\mu_{3}$ within the minimal-doubling range
to keep the two-flavor property irrespective of the way of use.
On the other hand, it is not clear whether we need the dimension-4 parameter tuning 
in the application to in-medium QCD. 
As shown in \cite{misumi}, for general values of $\mu_{3}$ and $d_4$, 
the dispersion relation in the Dirac operator is modified as
$D(p)\sim i\gamma_{i}p_{i}+ i\gamma_{4}p_{4} \sqrt{(1+d_{4})^{2}-\mu_{3}^{2}} +O(ap^{2})$.
It indicates that we may need to tune $d_{4}$ to correct the dispersion relation 
for given value of $\mu_{3}$ even if we consider the finite-density system.

\begin{figure}[t]
\begin{minipage}{0.5\hsize}
 \begin{center}
  \includegraphics[width=10em]{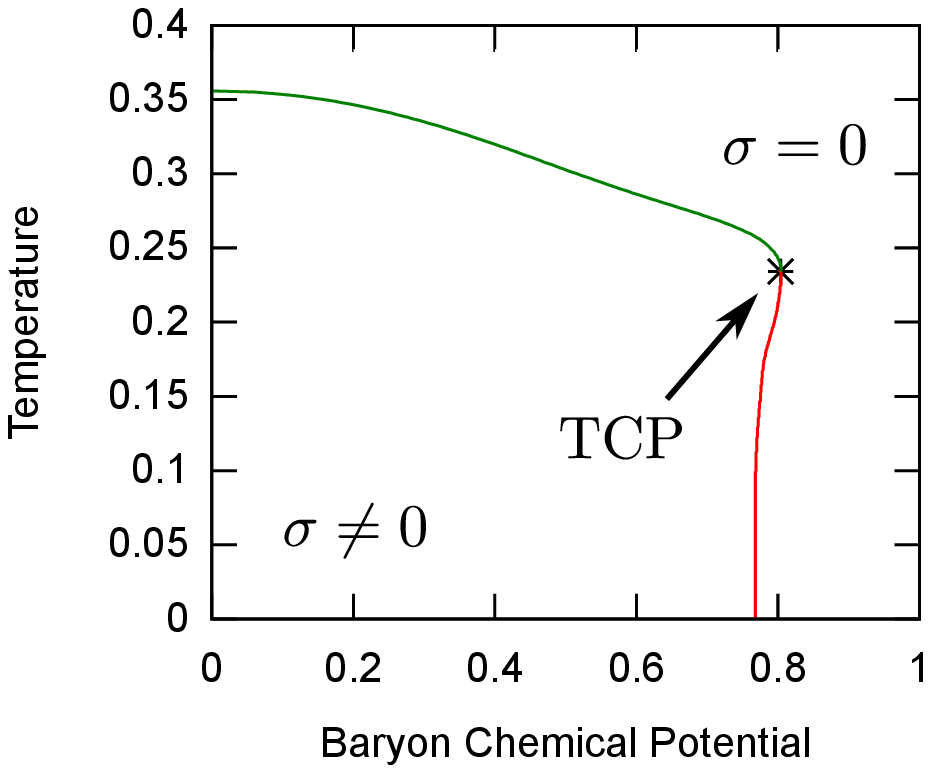}
 \end{center}
 \caption{Phase diagram for the chiral transition with $r=1$, $\mu_3=-0.9$
 and $m=0$ for $d_4=0$. 
 Green and red lines show 2nd and 1st transition lines, respectively.
 The transition order is changed from 2nd to 1st at the tricritical
 point $(\mu_B^{\rm tri},T^{\rm tri})=(0.804,0.234)$ \cite{MKO}.}
 \label{pb_r1}
\end{minipage}
\begin{minipage}{0.5\hsize}
 \begin{center}
   \includegraphics[width=7em]{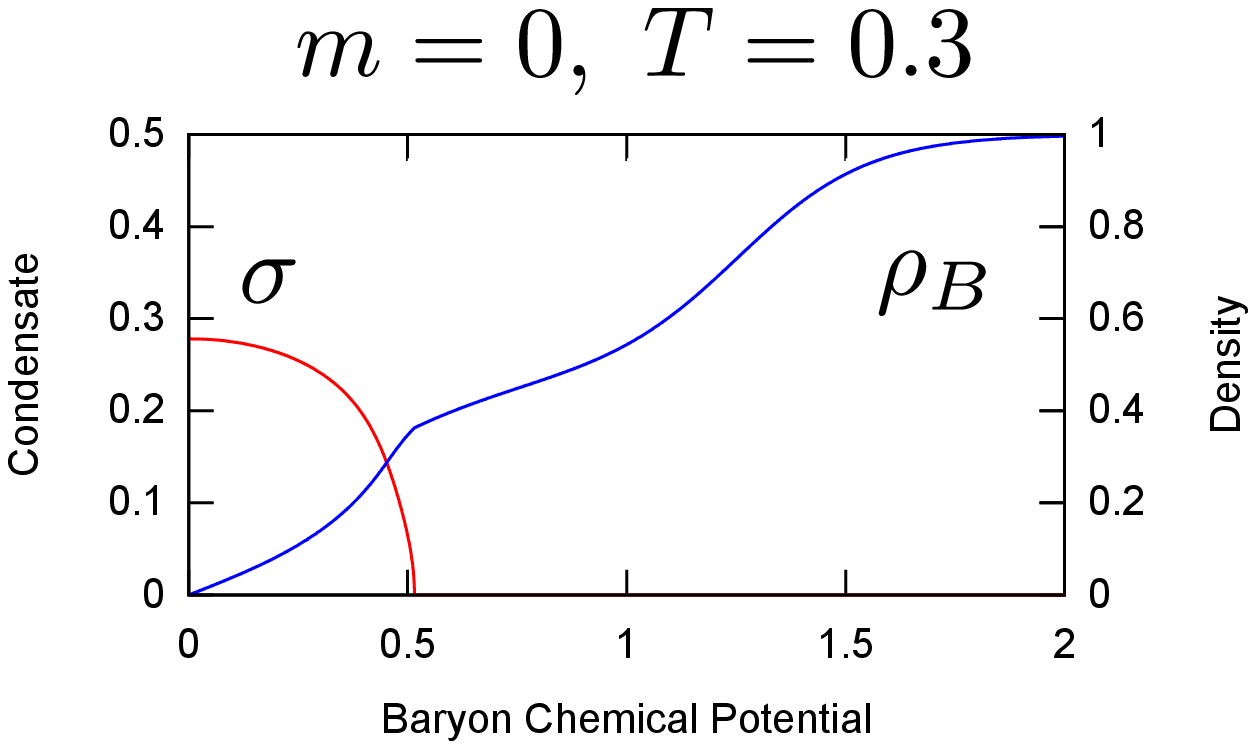} \quad
   \includegraphics[width=7em]{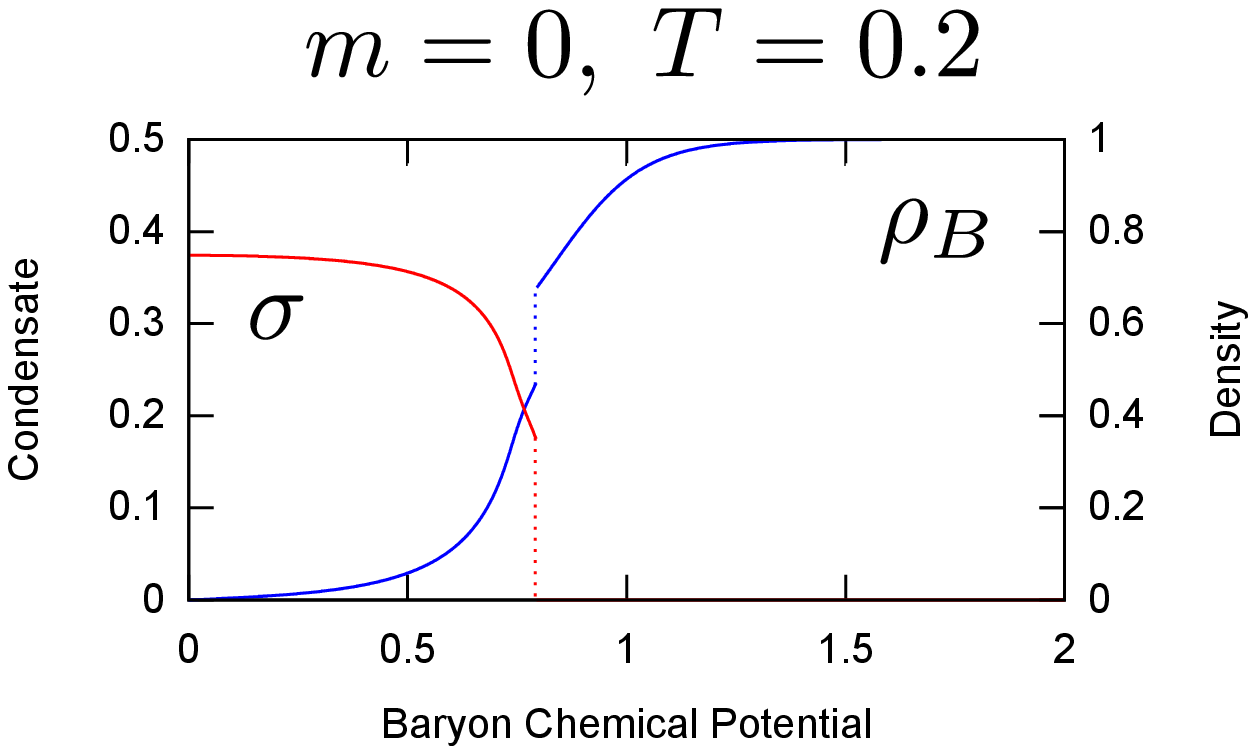} \\ \vspace{1.2em}
   \includegraphics[width=7em]{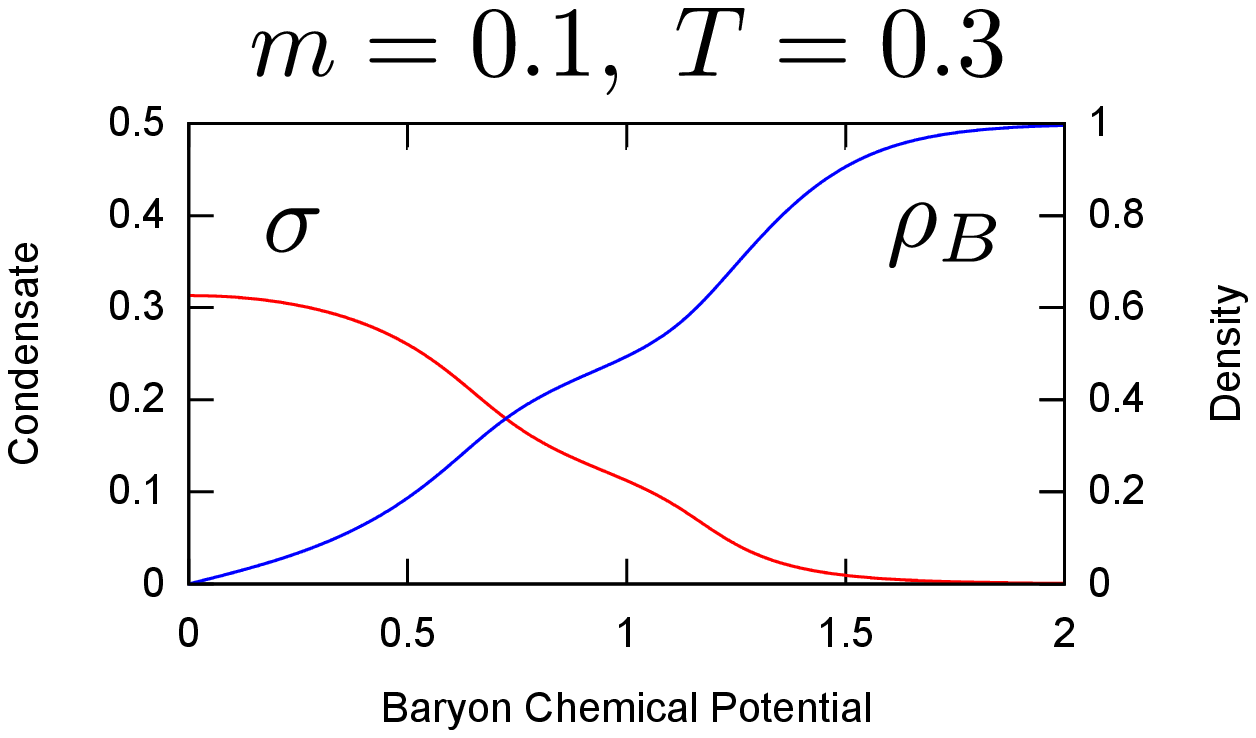} \quad
   \includegraphics[width=7em]{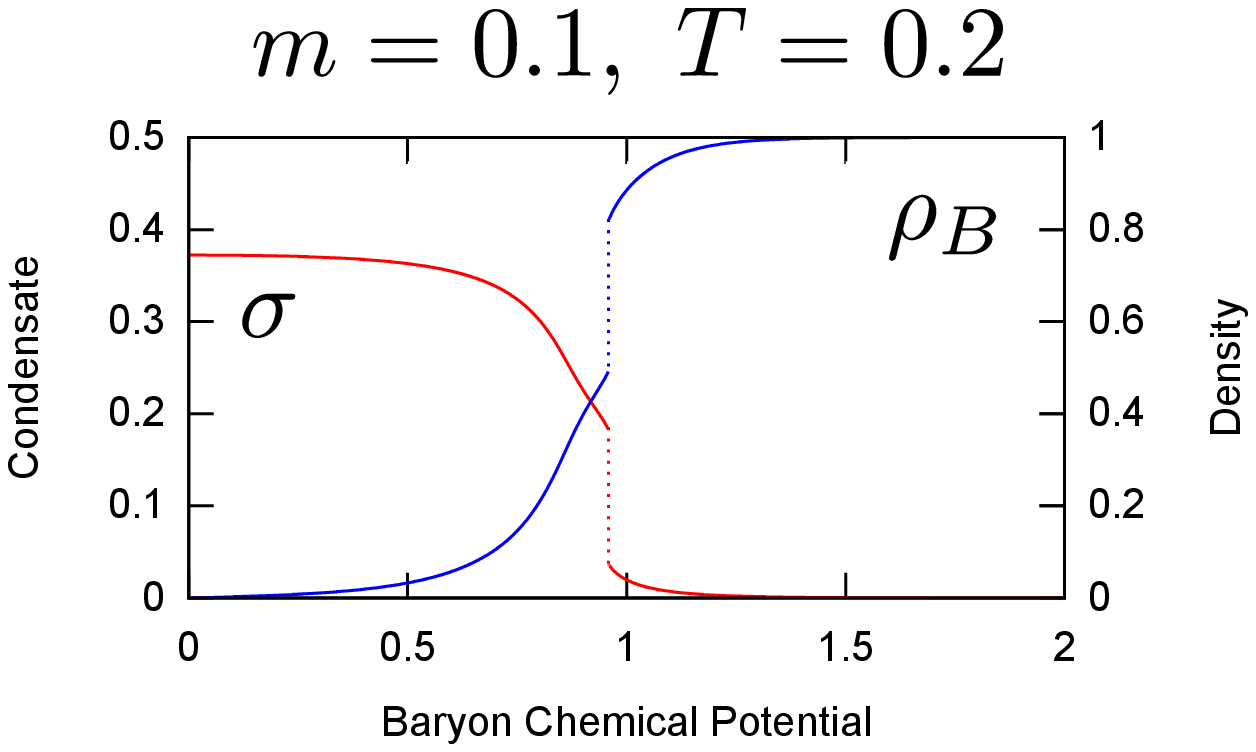}
  \end{center}
 \caption{Chiral condensate $\sigma$ and the baryon density $\rho_B$
 for (left) $T=0.3$ and (right) $T=0.2$ with $d_4=0$.
 Top and bottom panels show the massless $m=0$ case with 1st and 2nd phase transitions
  and massive $m=0.1$ case the crossover and 1st-order transitions \cite{MKO}.}
 \label{condensate_r1}
 \end{minipage}
\end{figure}

So far it is not clear how many parameters should be tuned in the application of
minimal-doubling fermions to the in-medium QCD.
However, Ref.~\cite{MKO} shows that the QCD phase diagram derived from this setup 
is consistent with the phenomenological predictions.
In Ref.~\cite{MKO}, the strong-coupling limit of lattice QCD with KW fermion in the presence of
temperature and density was studied.
We depict the results in Figs.~\ref{pb_r1} and \ref{condensate_r1}. 
The $(T,\mu_{B})$ phase diagram in Fig.~\ref{pb_r1} is consistent with 
strong-coupling lattice QCD with staggered fermions,
while there are some quantitative differences.
For example, the ratio of the transition baryon chemical potential at $T=0$
to the critical temperature at $\mu_B=0$ is given by
$R^{0}=\mu_c(T=0)/T_c(\mu_B=0)\sim 2.2$ for KW fermion while
$R^0_\mathrm{st} \simeq  \sim 1$ for staggered.
In the real world, this ratio is larger, $R^0 \sim 5.5$, and
larger $R^0$ with KW fermion in the strong coupling limit
may suggest smaller finite coupling corrections.
The location of the tricritical point in KW fermion is given by 
$R^\mathrm{tri}_\mathrm{KW} \simeq
3.4$ and while $R^\mathrm{tri}_\mathrm{st}= \simeq 2.0$
for staggered fermion.
It is consistent with the recent Monte-Carlo simulations,
which implies that the critical point does not exist
in the low baryon chemical potential region, $\mu_B/T \lesssim 3$.
These observations reveal the usefulness of KW fermion for research on 
QCD phase diagram. 

In the end of this section, we comment on the question whether the other 14 flavors 
are really decoupled in the continuum limit of minimal-doubling fermions.
If fermions with infinitely large chemical potential are not decoupled,
there should be some unphysical influence on the two-flavor QCD. 
This problem is discussed in \cite{Mslide}, and it is pointed out
that the subtraction of the free energy might be necessary to
derive physical thermodynamical quantities.
We still need further investigation in the application of this type of lattice fermions.


\section{Summary}
\label{sec:S}

In this review paper we discuss the recent progress of lattice fermions
and their possibilities of improving lattice simulations.
The formulations not only open a new course of lattice study,
but also give us feedback for the existing simulations. 

The flavored-mass terms, which are generalizations of the Wilson term,
produce new types of Wilson, domain-wall and overlap fermions.
The proper sum of the flavored-mass terms leads to the improved WIlson
fermions including the Brillouin fermion.
By introducing flavored mass into staggered fermions
we obtain staggered-Wilson fermions, which have the possibility
of reducing overlap numerical costs and staggered taste breaking. 
Although some numerical tests for these setups 
have not shown remarkable advantages, we expect the further tests 
with the improved versions could reveal their significance. 
The central branch in the flavored-mass fermions including Wilson fermion have the extra axial-type symmetry. This symmetry implies no necessity of fine-tuning of the mass parameter in
lattice QCD with this type. The lattice perturbation indicates that we have no additive mass renormalization in the interacting theory. From the strong-coupling lattice QCD it is shown that this extra symmetry is spontaneously broken with the associated NG boson emerging.
Although we need further investigation to judge its practical applicability, 
the interesting goal in this course is a two-flavor central-branch fermion, which 
can be a two-flavor setup with keeping chiral and hypercubic symmetry.
Although the minimal-doubling fermion is a chiral and ultra-local setup,   
we need to tune the several parameters to restore Lorentz and C, P, T 
invariances in the continuum limit.
This fermion can be seen as a lattice fermion with flavored chemical potential terms.
This interpretation gives us new and nonperturbative understanding on the tuning of the dimension-3 parameter based on the chiral phase structure.
Its use in the finite-density and -temperature systems was investigated
and the strong-coupling QCD shows that it produces the phase diagram
consistent to the phenomenological prediction. It requires numerical study 
to figure out requisite tuning process for this case.

\end{document}